\documentclass{article}
\usepackage{amsfonts}

\usepackage{graphicx}
\usepackage{amsmath}

\newtheorem{theorem}{Theorem}
\newtheorem{acknowledgement}[theorem]{Acknowledgement}

\newtheorem{remark}[theorem]{Remark}

\input{tcilatex}

\begin{document}

\title{Continuous measurement of canonical observables and limit Stochastic
Schr\"{o}dinger equations }
\author{John Gough\thanks{%
john.gough@ntu.ac.uk}, Andrei Sobolev\thanks{%
andrei.sobolev@ntu.ac.uk} \\
Department of Computing \& Mathematics\\
Nottingham-Trent University, Burton Street,\\
Nottingham NG1\ 4BU, United Kingdom.}
\date{\today}
\maketitle

\begin{abstract}
We derive the stochastic Schr\"{o}dinger equation for the limit of
continuous weak measurement where the observables monitored are canonical
position and momentum. To this end we extend an argument due to Smolianov
and Truman from the von Neumann model of \ indirect measurement of position
to the Arthurs and Kelly model for simultaneous measurement of position and
momentum. We only require unbiasedness of the detector states and an
integrability condition sufficient to ensure a central limit effect. Despite
taking a weak interaction, as opposed to weak measurement limit, \ the
resulting stochastic wave equation is of the same form as that derived in a
recent paper by Scott and Milburn for the specific case of joint Gaussian
states.
\end{abstract}

\section{Introduction}

The theory of continuous measurements of a quantum system, based on the
Ludwig formalism, was originally presented by Barchielli et al. \cite{BLP}.
Since then stochastic Schr\"{o}dinger equations describing the dynamical
evolution of the state of a system conditional on observations of a
monitored set of observables $\left\{ \hat{X}_{j}\right\} $ have been
developed by several authors \cite{GRW}\cite{Diosi}\cite{Belavkin2}\cite{GPR}
\cite{GP}. The generally accepted form is an adapted stochastic differential
equation for a vector state valued process $\left| \Psi \right\rangle $ of
the type (see, for instance \cite{JKupsch}) 
\begin{equation}
\left| d\Psi \right\rangle =\left\{ \frac{1}{i\hbar }\hat{H}-\sum_{j}\kappa
_{j}\left( \hat{X}_{j}-\left\langle \hat{X}_{j}\right\rangle \right)
^{2}\right\} \left| \Psi \right\rangle \,dt-\sum_{j}\sqrt{2\kappa _{j}}%
\left( \hat{X}_{j}-\left\langle \hat{X}_{j}\right\rangle \right) \left| \Psi
\right\rangle \,dB_{t}^{\left( j\right) }.  \label{SSE}
\end{equation}
where $\left\langle \hat{X}_{j}\right\rangle =\left\langle \Psi |\hat{X}%
_{j}\Psi \right\rangle $ are current observations (thus making the equation
non-linear) and $\left\{ B^{\left( j\right) }\right\} $ is a
multidimensional Wiener process. The constants $\kappa _{j}$ are positive
and $\hat{H}$ is the Hamiltonian governing the background evolution.

Note that the stochastic wave function is normalized, since, by the It\^{o}
rule $dB_{t}^{\left( j\right) }dB_{t}^{\left( k\right) }=\delta _{jk}dt$, 
\begin{equation}
d\left\| \Psi \right\| ^{2}=\left\langle d\Psi |\Psi \right\rangle
+\left\langle \Psi |d\Psi \right\rangle +\left\langle d\Psi |d\Psi
\right\rangle =0.
\end{equation}
If $\hat{Y}$ is an observable and if we set $\left\langle \hat{Y}%
\right\rangle =\left\langle \Psi |\,\hat{Y}\Psi \right\rangle $ then 
\begin{equation*}
d\left\langle \hat{Y}\right\rangle =\left\langle \mathcal{L}\left( \hat{Y}%
\right) \right\rangle dt-2\sum_{j}\sqrt{2\kappa _{j}}\,\text{cov}\left( \hat{%
Y},\hat{X}_{j}\right) \,dB_{t}^{\left( j\right) }
\end{equation*}
where $\mathcal{L}$ is the (self-dual) Lindblad generator 
\begin{equation*}
\mathcal{L}\left( \hat{Y}\right) =\frac{1}{i\hbar }\left[ \hat{Y},\hat{H}%
\right] +\sum_{j}\kappa _{j}\left\{ \left[ \hat{X}_{j},\hat{Y}\right] \hat{X}%
_{j}+\hat{X}_{j}\left[ \hat{Y},\hat{X}_{j}\right] \right\}
\end{equation*}
and we have the covariance 
\begin{equation*}
\text{cov}\left( \hat{Y},\hat{X}\right) =\frac{1}{2}\left\langle \hat{X}\hat{%
Y}+\hat{Y}\hat{X}\right\rangle -\left\langle \hat{X}\right\rangle
\left\langle \hat{Y}\right\rangle .
\end{equation*}

\bigskip

Stochastic Schr\"{o}dinger equations can alternatively be derived \cite
{Holevo} from quantum stochastic calculus \cite{HP}. Recently, a
probabilistic derivation based on limit theorems was presented by Smolianov
and Truman \cite{ST} which considered repeated position measurements
according to the von Neumann model. In this paper, we recap on the
calculations in \cite{ST} and correct some numerical errors. We then
generalize their argument to consider the simultaneous measurement of
position and momentum according to the Arthurs and Kelly model \cite{AK}:
this is non-trivial as we have to work with quantum rather than classical
probability distributions. This model has been examined by Scott and Milburn 
\cite{SM} with explicit calculations made for Gaussian states of the
measurement apparatus. They obtain the crucial result that the resulting
stochastic wave equation for continuous limit of weak measurements will be
of the form (\ref{SSE}) with $\hat{X}_{1}=\hat{q}$ and $\hat{X}_{2}=\hat{p}$%
. Our approach uses a different limit: a weak interactions limit. (Our
treatment of the Smolianov and Truman shows that the limit can be
interpreted as either as finite interactions with weak measurements (as they
consider) or as finite measurements with weak interactions: actually both
limits yield the same result.) We consider a general classes of states for
the apparatus where the only restriction is unbiasedness for the pointer
positions and momenta and a simple integrability condition, that the
constants $\kappa $ given by (\ref{kappa}) below are finite, which ensures a
central limit effect. The resulting stochastic wave equation is nevertheless
of the same form as that obtained by Scott and Milburn.

\subsection{Measurement Conditioned Wave Functions}

We consider a fixed Hilbert space $\frak{h}_{S}=L^{2}\left( \mathbb{R}%
^{n}\right) $ which describes our systems and consider indirect measurements
made on a second system (the apparatus) having Hilbert state space and $%
\frak{h}_{A}=L^{2}\left( \mathbb{R}^{m}\right) $. The joint space is $\frak{h%
}_{S}\otimes \frak{h}_{A}$. A vector state $\phi $ can be considered as a
function $\phi =\phi \left( \mathbf{q};\mathbf{q}^{\prime }\right) $ where $%
\mathbf{q} $ and $\mathbf{q}^{\prime }=\left( q_{1}^{\prime },\dots
,q_{m}^{\prime }\right) $ are appropriate coordinate variables for the
system and apparatus respectively. Let $\left( \hat{q}_{1}^{\prime },\dots ,%
\hat{q}_{m}^{\prime }\right) $ denote the corresponding position observables
for the apparatus: they form a maximal set of commuting, self-adjoint
operators on $\frak{h}_{A}$.

Suppose that the system and apparatus are initially prepared independently,
so that the joint state is the vector $\phi _{0}=\phi _{S}\otimes \phi _{A}$%
. Subsequently, the two undergo an interaction described by a unitary
operator $\hat{V}$. The state immediately prior to measurement is then $\phi
=\hat{V}\,\phi _{0}$. We then measure the observables $\left( \hat{q}%
_{1}^{\prime },\dots ,\hat{q}_{m}^{\prime }\right) $. According to the rules
of quantum mechanics, we obtain a family of classical random variables $%
\mathbf{Q}^{\prime }=\left( Q_{1}^{\prime },\dots ,Q_{m}^{\prime }\right) $
describing the measured values Here $Q_{j}^{\prime }$ is the random variable
describing the measured value of the observable $\hat{q}_{j}^{\prime }$.
distributed according to 
\begin{eqnarray}
\mathbb{E}\left[ \exp \left\{ i\sum_{j=1}^{m}t_{j}Q_{j}^{\prime }\right\} %
\right] &\equiv &\left\langle \phi |\exp \left\{ i\sum_{j=1}^{m}t_{j}\hat{q}%
_{j}^{\prime }\right\} \phi \right\rangle  \notag \\
&\equiv &\int_{\mathbb{R}^{m}}\rho \left( \mathbf{q}^{\prime }\right) \exp
\left\{ i\sum_{j=1}^{m}t_{j}q_{j}^{\prime }\right\} d^{m}q^{\prime },
\end{eqnarray}
where $\rho \left( \mathbf{q}^{\prime }\right) =\int \left| \left\langle 
\mathbf{q;q}^{\prime }|\,\phi \right\rangle \right| ^{2}\,d^{n}q$.

We then consider the function 
\begin{equation}
\psi \left( \mathbf{q}|\mathbf{q}^{\prime }\right) :=\frac{\left\langle 
\mathbf{q};\mathbf{q}^{\prime }|\phi \right\rangle }{\sqrt{\rho \left( 
\mathbf{q}^{\prime }\right) }}
\end{equation}
and introduce the \textit{stochastic wave-function} $\Psi $ defined by

\begin{equation}
\left\langle \mathbf{q}|\Psi \right\rangle =\psi \left( \mathbf{q}|\mathbf{Q}%
^{\prime }\right)
\end{equation}
If we do not wish to make the $\mathbf{q}$-dependence explicit, we shall
write $\Psi =\Psi _{\mathbf{Q}^{\prime }}$. The stochastic wave-function
interpretation comes about as the mathematical re-interpretation of the
wave-function as the $\frak{h}_{S}$-valued random variable dependent on the
measured random variables $\mathbf{Q}^{\prime }=\left( Q_{1}^{\prime },\dots
,Q_{m}^{\prime }\right) $.

Let $\hat{X}$ is an observable of the system, only. We set $\bar{X}\left( 
\mathbf{Q}^{\prime }\right) =\left\langle \Psi _{\mathbf{Q}^{\prime }}|\hat{X%
}\,\Psi _{\mathbf{Q}^{\prime }}\right\rangle =\int \psi ^{\ast }\left( 
\mathbf{q}|\mathbf{Q}^{\prime }\right) \left( \hat{X}\psi \right) \left( 
\mathbf{q}|\mathbf{Q}^{\prime }\right) \,d^{n}q$. Then the average value of
the random variable $\bar{X}\left( \mathbf{Q}^{\prime }\right) $ will be 
\begin{equation}
\mathbb{E}\left[ \bar{X}\left( \mathbf{Q}^{\prime }\right) \right] =\int 
\bar{X}\left( \mathbf{q}^{\prime }\right) \,\rho \left( \mathbf{q}^{\prime
}\right) \,d^{m}q^{\prime }
\end{equation}
and this is clearly $\left\langle \phi |\hat{X}\,\phi \right\rangle =\int
\phi ^{\ast }\left( \mathbf{q};\mathbf{q}^{\prime }\right) \left( \hat{X}%
\phi \right) \left( \mathbf{q};\mathbf{q}^{\prime }\right)
\,d^{n}qd^{m}q^{\prime }$.

\subsection{Repeated Measurements}

Next suppose that several measurements are made at regular intervals of time 
$\tau $. Let $A\left( j\right) $ denote the apparatus employed for the $j$%
-th measurement and let $\phi _{A\left( j\right) }$ be its state prior to
interaction with the system. (We may either consider an assembly of
apparatuses $A\left( 1\right) ,A\left( 2\right) ,A\left( 3\right) ,\cdots $
or a single apparatus whose state is forced to be $\phi _{A\left( j\right) }$
somehow just before the $j$-th measurement for each $j$.) Between
measurements, the system is allowed to evolve according to the Hamiltonian $%
\hat{H}$ on $\frak{h}_{S}$.

If we denote the conditioned wave-function immediately after the $n$-th
measurement by $\psi _{n}=\psi _{n}\left( x|\mathbf{Q}_{1}^{\prime };\cdots ,%
\mathbf{Q}_{n}^{\prime }\right) $, then we have the iterative relation 
\begin{equation}
\psi _{n}\left( \mathbf{q}|\mathbf{q}_{1}^{\prime };\cdots ;\mathbf{q}%
_{n}^{\prime }\right) =\frac{1}{\sqrt{\rho \left( \mathbf{q}_{n}^{\prime
}\right) }}\;\left\langle \mathbf{q};\mathbf{q}_{n}^{\prime }|\,e^{\tau \hat{%
H}/i\hbar }\hat{V}_{n}\,\psi _{n-1}\otimes \phi _{A\left( n\right)
}\right\rangle .
\end{equation}

\section{Measurement of Position}

We begin with the simplest problem of monitoring a single observable. We
take this to be a position coordinate $\hat{q}$.

As system-apparatus interaction, we take a linear coupling of the system
position with the apparatus momentum $\hat{p}^{\prime }$. The interaction is
then given by the unitary $\hat{V}=\exp \left\{ \mu \hat{q}\hat{p}^{\prime
}/ih\right\} $. We then measure the position observable, $\hat{q}^{\prime }$%
, for the apparatus. This is essentially the von Neumann model for indirect
measurements. The apparatus state prior to measurement is taken to have a
real symmetric wave-function and so its mean position and moment are zero.

The calculations in this section follow the same arguments as in Smolyanov
and Truman \cite{ST}.

\subsection{Von Neumann's Model}

The interaction between the system and apparatus is then given by 
\begin{equation}
\phi _{S}\left( q\right) \phi _{A}\left( q^{\prime }\right) \mapsto \phi
_{S}\left( q\right) \phi _{A}\left( q^{\prime }-\mu q\right) .
\end{equation}
The probability density for the apparatus position after interaction is then 
\begin{equation}
\rho \left( q^{\prime }\right) =\int \left| \phi _{S}\left( q\right) \right|
^{2}\left| \phi _{A}\left( q^{\prime }-\mu q\right) \right| ^{2}dq
\end{equation}
which is a convolution of two probability densities.

We choose $\phi _{A}$ to have the form 
\begin{equation}
\phi _{A}\left( q^{\prime }\right) =\frac{1}{\sqrt{\sigma }}\chi \left( 
\frac{\left| q^{\prime }\right| ^{2}}{\sigma ^{2}}\right)  \label{CHI}
\end{equation}
where $\chi $ is real-valued and normalized so that $\int_{-\infty }^{\infty
}\chi \left( y^{2}\right) dy=1=\int_{-\infty }^{\infty }\chi \left(
y^{2}\right) y^{2}dy$. In particular, $Y=\frac{1}{\sigma }Q_{0}^{\prime }$
will be a mean-zero, unit-variance random variable. We then have the
decomposition 
\begin{equation}
Q^{\prime }=\sigma Y+\mu \left( Z+\left\langle \hat{q}\right\rangle \right)
\end{equation}
where $Z=Q_{S}-\left\langle \hat{q}\right\rangle $ is centered.

Assuming that $\chi $ is analytic, we get an expansion of the form 
\begin{equation*}
\phi _{A}\left( Q^{\prime }-\mu q\right) =\frac{1}{\sqrt{\sigma }}\chi
\left( Y^{2}\right) \left\{ 1+\frac{\chi ^{\prime }\left( Y^{2}\right) }{%
\chi \left( Y^{2}\right) }\varepsilon +\frac{1}{2!}\frac{\chi ^{\prime
\prime }\left( Y^{2}\right) }{\chi \left( Y^{2}\right) }\varepsilon
^{2}+\cdots \right\}
\end{equation*}
where $\varepsilon =\sigma ^{-2}\left| Q^{\prime }-\mu q\right| ^{2}-Y^{2}$.

Likewise the factor $\rho \left( Q^{\prime }\right) ^{-1/2}$ can be
expanded. First of all we observe that 
\begin{eqnarray*}
\int \left| \phi _{S}\left( q\right) \right| ^{2}\varepsilon dq &=&2\frac{%
\mu }{\sigma }ZY+\left( \frac{\mu }{\sigma }\right) ^{2}\left( Z^{2}+\sigma
_{q}^{2}\right) \\
\int \left| \phi _{S}\left( q\right) \right| ^{2}\varepsilon ^{2}dq
&=&4\left( \frac{\mu }{\sigma }\right) ^{2}\left( Z^{2}+\sigma
_{q}^{2}\right) Y^{2}+O\left( \left( \frac{\mu }{\sigma }\right) ^{3}\right)
\end{eqnarray*}
where we set $\sigma _{q}^{2}:=\left\langle \hat{q}^{2}\right\rangle
-\left\langle \hat{q}\right\rangle ^{2}$ (the variance of the observable $%
\hat{q}$ for state $\phi _{S}$) and treat $\left( \frac{\mu }{\sigma }%
\right) $ as small parameter. 
\begin{align*}
\rho \left( Q^{\prime }\right) & =\int \left| \phi _{S}\left( q\right)
\right| ^{2}\frac{1}{\sigma ^{2}}\chi ^{2}\left( \frac{\left| Q^{\prime
}-\mu q\right| ^{2}}{\sigma ^{2}}\right) dq \\
& =\int \left| \phi _{S}\left( q\right) \right| ^{2}\frac{\chi ^{2}\left(
Y^{2}\right) }{\sigma ^{2}}\left\{ 1+\frac{2\chi ^{\prime }\left(
Y^{2}\right) }{\chi \left( Y^{2}\right) }\varepsilon +\frac{1}{2!}\left[ 
\frac{\chi ^{\prime }\left( Y^{2}\right) ^{2}}{\chi \left( Y^{2}\right) ^{2}}%
+\frac{\chi ^{\prime \prime }\left( Y^{2}\right) }{\chi \left( Y^{2}\right) }%
\right] \varepsilon ^{2}+\cdots \right\} dq \\
& =\frac{\chi ^{2}\left( Y^{2}\right) }{\sigma ^{2}}\left\{ 1+\frac{4\chi
^{\prime }\left( Y^{2}\right) }{\chi \left( Y^{2}\right) }\frac{\mu }{\sigma 
}ZY\right. \\
& \left. +2\frac{\mu ^{2}}{\sigma ^{2}}\left[ \frac{\chi ^{\prime }\left(
Y^{2}\right) }{\chi \left( Y^{2}\right) }+2\frac{\chi ^{\prime }\left(
Y^{2}\right) ^{2}}{\chi \left( Y^{2}\right) ^{2}}+2\frac{\chi ^{\prime
\prime }\left( Y^{2}\right) }{\chi \left( Y^{2}\right) }\right] \left(
Z^{2}+\sigma _{q}^{2}\right) Y^{2}+O\left( \left( \frac{\mu }{\sigma }%
\right) ^{3}\right) \right\}
\end{align*}
We obtain to lowest orders 
\begin{gather}
\psi \left( q|Q^{\prime }\right) =\left\{ 1-2\frac{\mu }{\sigma }\frac{\chi
^{\prime }}{\chi }\left( q-\left\langle \hat{q}\right\rangle \right)
Y+\right.  \notag \\
+\frac{\mu ^{2}}{\sigma ^{2}}\left[ \left( \frac{\chi ^{\prime }}{\chi }+2%
\frac{\chi ^{\prime \prime }}{\chi }Y^{2}\right) \left( q-\left\langle \hat{q%
}\right\rangle \right) ^{2}-\left( \frac{\chi ^{\prime }}{\chi }+2\frac{\chi
^{\prime \prime }}{\chi }Y^{2}+2\left( \frac{\chi ^{\prime }}{\chi }\right)
^{2}Y^{2}\right) \sigma _{q}^{2}\right.  \notag \\
\left. \left. +2Z\left( 2\left( \frac{\chi ^{\prime }}{\chi }\right)
^{2}Y^{2}-\frac{\chi ^{\prime }}{\chi }-2\frac{\chi ^{\prime \prime }}{\chi }%
Y^{2}\right) \left( q-\left\langle \hat{q}\right\rangle \right) \right]
+O\left( \left( \frac{\mu }{\sigma }\right) ^{3}\right) \right\} \phi
_{S}\left( q\right) .
\end{gather}
where $\chi ,\chi ^{\prime }$ and $\chi ^{\prime \prime }$ are evaluated at $%
Y^{2}$. Note that the terms involving $Z^{2}$ cancel to this order.

\subsection{Continuous Measurement Limit}

We now consider sequential measurements with the successive apparatus states
prior to interaction being copies of $\phi _{A}$ chosen as above (\ref{CHI}).

Adopting the scaling $\left( \frac{\mu }{\sigma }\right) =\sqrt{\tau }$, we
get 
\begin{gather*}
\frac{1}{\tau }[\psi _{n}-\psi _{n-1}]=\left\{ \frac{1}{i\hbar }\hat{H}-2%
\sqrt{\frac{1}{\tau }}\frac{\chi _{n}^{\prime }}{\chi _{n}}\left(
q-\left\langle \hat{q}\right\rangle \right) Y_{n}\right. \\
+\left[ \left( \frac{\chi _{n}^{\prime }}{\chi _{n}}+2\frac{\chi
_{n}^{\prime \prime }}{\chi _{n}}Y_{n}^{2}\right) \left( q-\left\langle \hat{%
q}\right\rangle \right) ^{2}-\left( \frac{\chi _{n}^{\prime }}{\chi _{n}}+2%
\frac{\chi _{n}^{\prime \prime }}{\chi _{n}}Y_{n}^{2}+2\left( \frac{\chi
_{n}^{\prime }}{\chi _{n}}\right) ^{2}Y_{n}^{2}\right) \sigma _{q}^{2}\right.
\\
\left. \left. +2Z_{n}\left( 2\left( \frac{\chi _{n}^{\prime }}{\chi _{n}}%
\right) ^{2}Y_{n}^{2}-\frac{\chi _{n}^{\prime }}{\chi _{n}}-2\frac{\chi
_{n}^{\prime \prime }}{\chi _{n}}Y_{n}^{2}\right) \left( q-\left\langle \hat{%
q}\right\rangle \right) \right] +O\left( \tau ^{1/2}\right) \right\} \psi
_{n-1}.
\end{gather*}
where $\chi _{n}=\chi \left( Y_{n}^{2}\right) $, etc.

In the limit $\tau \rightarrow 0$ we expect from the law of large numbers
that 
\begin{eqnarray*}
\lim_{\tau \rightarrow 0}\tau \sum_{j=1}^{\left[ t/\tau \right] }\left( 
\frac{\chi _{j}^{\prime }}{\chi _{j}}+2\frac{\chi _{j}^{\prime \prime }}{%
\chi _{j}}Y_{j}^{2}\right) &=&-\kappa t; \\
\lim_{\tau \rightarrow 0}\tau \sum_{j=1}^{\left[ t/\tau \right] }f\left(
Y_{j}^{2}\right) Z_{j} &=&0;
\end{eqnarray*}
for suitable $f$. Here $\kappa =-\int_{-\infty }^{\infty }\left( \frac{\chi
^{\prime }}{\chi }+2\frac{\chi ^{\prime \prime }}{\chi }y^{2}\right) \chi
^{2}dy$, after some algebra and an integration by parts we have that 
\begin{equation}
\kappa =-\int_{-\infty }^{\infty }\left( \chi ^{\prime }\chi +2\chi \chi
^{\prime \prime }y^{2}\right) dy=2\int_{-\infty }^{\infty }\left( \chi
^{\prime }y\right) ^{2}dy>0.  \label{kappa}
\end{equation}
Inspecting the coefficient of $\sigma _{q}^{2}$, we are lead to consider 
\begin{equation*}
\lim_{\tau \rightarrow 0}\tau \sum_{j=1}^{\left[ t/\tau \right] }\left( 
\frac{\chi _{j}^{\prime }}{\chi _{j}}+2\frac{\chi _{j}^{\prime \prime }}{%
\chi _{j}}Y_{j}^{2}+2\left( \frac{\chi _{j}^{\prime }}{\chi _{j}}\right)
^{2}Y_{j}^{2}\right) =\theta t;
\end{equation*}
where $\theta =\int_{-\infty }^{\infty }\left( \chi \chi ^{\prime }+2\chi
^{\prime 2}y^{2}+2\chi ^{\prime \prime }\chi y^{2}\right) dy$ and comparison
with the above integration by parts shows $\theta =0.$

Likewise, from the central limit theorem, we expect that 
\begin{equation*}
\lim_{\tau \rightarrow 0}\frac{1}{\sqrt{\tau }}\sum_{j=1}^{\left[ t/\tau %
\right] }\frac{\chi ^{\prime }\left( Y_{j}^{2}\right) }{\chi \left(
Y_{j}^{2}\right) }Y_{j}=\sqrt{\frac{\kappa }{2}}B_{t}
\end{equation*}
where $B_{t}$ is a standard Wiener process.

\bigskip

We therefore are lead to the stochastic differential equation 
\begin{equation}
\left| d\Psi \right\rangle =\left\{ \frac{1}{i\hbar }\hat{H}-\kappa \left( 
\hat{q}-\left\langle \hat{q}\right\rangle \right) ^{2}\right\} \left| \Psi
\right\rangle \,dt-\sqrt{2\kappa }\left( \hat{q}-\left\langle \hat{q}%
\right\rangle \right) \left| \Psi \right\rangle \,dB_{t}.  \label{SSEq}
\end{equation}

\begin{remark}
The limit $\frac{\mu }{\sigma }\rightarrow 0$ can occur in two distinct
ways. We can consider a weak coupling limit $\mu \rightarrow 0$ where the
measurement process is fixed $\left( \sigma =1\right) $. Alternatively we
can consider a weak measurement limit $\sigma \rightarrow \infty $ where the
interaction is fixed $\left( \mu =1\right) $.
\end{remark}

\begin{remark}
The equation (\ref{SSEq}) can clearly be generalized to (\ref{SSE}) provide
that we consider indirect measurement of commuting system observables.
\end{remark}

\begin{remark}
The result (\ref{SSEq}) corrects some numerical errors made in \cite{ST}
with regard to the coefficients.
\end{remark}

\section{Measurement of Position \& Momentum}

We take the system to have canonical position and momentum observables $\hat{%
q}$ and $\hat{p}$, respectively. Our apparatus consist of two distinct
components, $A^{\prime }$ and $A^{\prime \prime }$, which have canonical
observables $\hat{q}^{\prime },\hat{p}^{\prime }$ and $\hat{q}^{\prime
\prime },\hat{p}^{\prime \prime }$, respectively. Our aim is to couple the
system to the apparatus by interaction and make inferences about the system
based on simultaneous measurements of the position of $A^{\prime }$ (that
is, $\hat{q}^{\prime }$) and the momentum of $A^{\prime \prime }$ (that is, $%
\hat{p}^{\prime \prime }$).

Initially we take the system to be prepared in state $\phi _{S}$ while the
apparatus is prepared in a state $\phi _{A}\in \frak{h}_{A}=\frak{h}%
_{A^{\prime }}\otimes \frak{h}_{A^{\prime \prime }}$. We assume that the two
components of the apparatus are not entangled - that is, $\phi _{A}\left(
q^{\prime },p^{\prime \prime }\right) =\phi _{A^{\prime }}\left( q^{\prime
}\right) \phi _{A^{\prime \prime }}\left( p^{\prime \prime }\right) $.

\bigskip

The interaction between the system and apparatus is described by the unitary
operator 
\begin{equation}
\hat{V}=\exp \left\{ \left( \mu \hat{p}^{\prime }\hat{q}-\nu \hat{q}^{\prime
\prime }\hat{p}\right) /i\hbar \right\}
\end{equation}
and we define an operator $\mathcal{\hat{V}}\left( q^{\prime },p^{\prime
\prime }\right) $ on $\frak{h}_{S}$\ by 
\begin{equation}
\left( \mathcal{\hat{V}}\left( q^{\prime },p^{\prime \prime }\right) \phi
_{S}\right) \left( q\right) =\left\langle q,q^{\prime },p^{\prime \prime }|%
\hat{V}\;\phi _{S}\otimes \phi _{A}\right\rangle .
\end{equation}
We see that 
\begin{eqnarray*}
\mathcal{\hat{V}}\left( q^{\prime },p^{\prime \prime }\right)
&=&\int_{\Gamma \times \Gamma }\frac{dq^{\prime \prime }dp^{\prime }}{2\pi
\hbar }\frac{d\bar{q}^{\prime }d\bar{p}^{\prime \prime }}{2\pi \hbar }%
\left\langle q^{\prime }|p^{\prime }\right\rangle \left\langle p^{\prime
\prime }|q^{\prime \prime }\right\rangle \hat{W}\left( -\nu q^{\prime \prime
},-\mu p^{\prime }\right) \left\langle p^{\prime }|\bar{q}^{\prime
}\right\rangle \left\langle q^{\prime \prime }|\bar{p}^{\prime \prime
}\right\rangle \left\langle \bar{q}^{\prime },\bar{p}^{\prime \prime }|\phi
_{A}\right\rangle \\
&=&\int_{\Gamma \times \Gamma }\frac{dqdp}{2\pi \hbar }\frac{d\bar{q}d\bar{p}%
}{2\pi \hbar }\;e^{\left( p\bar{q}-q\bar{p}\right) /i\hbar }\,\left\langle
q^{\prime }-\mu \bar{q},p^{\prime \prime }-\nu \bar{p}|\phi
_{A}\right\rangle \;\hat{W}\left( q,p\right)
\end{eqnarray*}
and so we obtain $\mathcal{\hat{V}}\left( q^{\prime },p^{\prime \prime
}\right) $ as a Weyl-quantized operator (see appendix) 
\begin{equation}
\mathcal{\hat{V}}\left( q^{\prime },p^{\prime \prime }\right) \equiv \left[
\left\langle q^{\prime }-\mu \hat{q},p^{\prime \prime }-\nu \hat{p}|\phi
_{A}\right\rangle \right] _{\text{Weyl}}.
\end{equation}

\bigskip

We next denote the random variables obtained by observing $\hat{q}^{\prime }$
and $\hat{p}^{\prime \prime }$\ by $Q^{\prime }$ and $P^{\prime \prime }$
respectively. As the observables commute, we are able to assign a joint
probability to $Q^{\prime },P^{\prime \prime }$ once a density matrix is
prescribed.

A stochastic wave-function dependent on the observed position and momentum
variables of the apparatus is 
\begin{equation}
\Psi =\frac{\mathcal{\hat{V}}\left( Q^{\prime },P^{\prime \prime }\right) }{%
\sqrt{\rho \left( Q^{\prime },P^{\prime \prime }\right) }}\phi _{S}
\end{equation}
where $\rho \left( q^{\prime },p^{\prime \prime }\right) =\int \left|
\left\langle q,q^{\prime },p^{\prime \prime }|\,\hat{V}\phi \right\rangle
\right| ^{2}dq\equiv \left\| \mathcal{\hat{V}}\left( q^{\prime },p^{\prime
\prime }\right) \phi _{S}\right\| ^{2}$.

\bigskip

Under the action of the unitary $\hat{V}$ we have the relations 
\begin{equation}
\hat{V}^{\dagger }\,\hat{q}^{\prime }\,\hat{V}=\hat{q}^{\prime }+\mu \hat{q}-%
\frac{1}{2}\mu \nu \hat{q}^{\prime \prime };\quad \hat{V}^{\dagger }\,\hat{p}%
^{\prime \prime }\,\hat{V}=\hat{p}^{\prime \prime }+\nu \hat{p}-\frac{1}{2}%
\mu \nu \hat{p}^{\prime }.
\end{equation}
(The action of $\hat{V}$ can be understood as that due to the evolution in
unit time governed by Hamiltonian $\mu \hat{p}^{\prime }\hat{q}-\nu \hat{q}%
^{\prime \prime }\hat{p}$. This Hamiltonian generates a linear set of
equation for the canonical variables which is readily soluble. The variables 
$\hat{p}^{\prime }$ and $\hat{q}^{\prime \prime }$ are evidently invariants.)

\bigskip

The joint probability distribution of $Q^{\prime }$ and $P^{\prime \prime }$
is determined from the characteristic function 
\begin{equation*}
\mathbb{E}\left[ e^{i\left( \alpha Q^{\prime }+\beta P^{\prime \prime
}\right) }\right] =\left\langle \phi |\,e^{i\left( \alpha \hat{q}^{\prime
}+\beta \hat{p}^{\prime \prime }\right) }\phi \right\rangle
\end{equation*}
where $\phi =\hat{V}\left( \phi _{S}\otimes \phi _{A}\right) $ is the
post-interaction state. Using the lemma, we can write the characteristic
function as 
\begin{equation*}
\left\langle \phi _{S}|\,e^{i\left( \alpha \mu \hat{q}+\beta \nu \hat{p}%
\right) }\phi _{S}\right\rangle \times \left\langle \phi _{A^{\prime
}}|\,e^{i\left( \alpha \hat{q}^{\prime }-\frac{1}{2}\beta \mu \nu \hat{p}%
^{\prime }\right) }\phi _{A^{\prime }}\right\rangle \times \left\langle \phi
_{A^{\prime \prime }}|\,e^{i\left( \beta \hat{p}^{\prime \prime }-\frac{1}{2}%
\alpha \mu \nu \hat{q}^{\prime \prime }\right) }\phi _{A^{\prime \prime
}}\right\rangle
\end{equation*}
The result is that we have the following decompositions into sums of
independent random variables: 
\begin{equation}
Q^{\prime }=Q_{0}^{\prime }+\mu Q_{S}-\frac{1}{2}\mu \nu Q_{0}^{\prime
\prime };\quad P^{\prime \prime }=P_{0}^{\prime \prime }+\nu P_{S}-\frac{1}{2%
}\mu \nu P_{0}^{\prime }.  \label{decomp}
\end{equation}
Specifically, 
\begin{eqnarray*}
\Pr \left[ q\leq Q_{S}<q+dq\right] &=&\left| \phi _{S}\left( q\right)
\right| ^{2}dq \\
\Pr \left[ q^{\prime }\leq Q_{0}^{\prime }<q^{\prime }+dq^{\prime }\right]
&=&\left| \phi _{A^{\prime }}\left( q^{\prime }\right) \right|
^{2}dq^{\prime } \\
\Pr \left[ q^{\prime \prime }\leq Q_{0}^{\prime \prime }<q^{\prime \prime
}+dq^{\prime \prime }\right] &=&\left| \left\langle q^{\prime \prime }|\phi
_{A^{\prime \prime }}\right\rangle \right| ^{2}dq^{\prime \prime }
\end{eqnarray*}
while a similar set of laws hold for the $P$'s.

\bigskip

\begin{remark}
Some caution is needed here: a joint distribution for $Q^{\prime },P^{\prime
\prime }$ exists as they correspond to commuting observables, however this
is not the case for any of the pairs $Q_{S},P_{S}$\ or $Q_{0}^{\prime
},P_{0}^{\prime }$\ or $Q_{0}^{\prime \prime },P_{0}^{\prime \prime }$.
\end{remark}

\begin{remark}
The parameters $\mu ,\nu $ are free. In \cite{SM} they are chosen as $\nu
=\mu ^{-1}$ so that symplectic area is preserved: the single parameter $\mu $
is referred to therein as a squeezing parameter .
\end{remark}

\subsection{Perturbative Expansions}

We take the initial states of the apparatus to be 
\begin{equation*}
\phi _{A^{\prime }}\left( q^{\prime }\right) =\chi \left( \left| q^{\prime
}\right| ^{2}\right) ,\quad \phi _{A^{\prime \prime }}\left( p^{\prime
\prime }\right) =\Lambda \left( \left| p^{\prime \prime }\right| ^{2}\right)
\end{equation*}
Here we shall adopt the weak coupling scheme and fix $\sigma =1$ while $\mu $
and $\nu $ are the small parameters. Expanding to lowest orders, we see 
\begin{gather*}
\left\langle q^{\prime }-\mu q,p^{\prime \prime }-\nu p|\phi
_{A}\right\rangle =\chi \left( \left| q^{\prime }-\mu q\right| ^{2}\right)
\Lambda \left( \left| p^{\prime \prime }-\nu p\right| ^{2}\right) \\
=\chi \Lambda \left\{ 1+\frac{\chi ^{\prime }}{\chi }\varepsilon _{q}+\frac{1%
}{2!}\frac{\chi ^{\prime \prime }}{\chi }\varepsilon _{q}^{2}+\cdots
\right\} \left\{ 1+\frac{\Lambda ^{\prime }}{\Lambda }\varepsilon _{p}+\frac{%
1}{2!}\frac{\Lambda ^{\prime \prime }}{\Lambda }\varepsilon _{p}^{2}+\cdots
\right\}
\end{gather*}
where $\chi =\chi \left( \left| q^{\prime }\right| ^{2}\right) $ and $%
\Lambda =\Lambda \left( \left| p^{\prime \prime }\right| ^{2}\right) $,
etc., and 
\begin{equation*}
\varepsilon _{q}=-2\mu q^{\prime }q+\mu ^{2}q^{2},\;\varepsilon _{p}=-2\nu
p^{\prime \prime }p+\nu ^{2}p^{2}.
\end{equation*}
Taking Weyl quantization yields 
\begin{gather*}
\mathcal{\hat{V}}\left( q^{\prime },p^{\prime \prime }\right) =\chi \Lambda
\left\{ 1-2\mu \frac{\chi ^{\prime }}{\chi }q^{\prime }\hat{q}-2\nu \frac{%
\Lambda ^{\prime }}{\Lambda }p^{\prime \prime }\hat{p}\right. \\
\left. +\mu ^{2}\left( \frac{\chi ^{\prime }}{\chi }+2\frac{\chi ^{\prime
\prime }}{\chi }q^{\prime 2}\right) \hat{q}^{2}+\nu ^{2}\left( \frac{\Lambda
^{\prime }}{\Lambda }+2\frac{\Lambda ^{\prime \prime }}{\Lambda }p^{\prime
\prime 2}\right) \hat{p}^{2}+2\mu \nu \frac{\chi ^{\prime }}{\chi }\frac{%
\Lambda ^{\prime }}{\Lambda }q^{\prime }p^{\prime \prime }\left( \hat{q}\hat{%
p}+\hat{p}\hat{q}\right) +\cdots \right\}
\end{gather*}
and the joint probability density for the observations is 
\begin{gather*}
\rho \left( q^{\prime },p^{\prime \prime }\right) =\left\langle \phi _{S}|\,%
\mathcal{\hat{V}}\left( q^{\prime },p^{\prime \prime }\right) ^{2}\phi
_{S}\right\rangle \\
=\chi ^{2}\Lambda ^{2}\left\{ 1-4\mu \frac{\chi ^{\prime }}{\chi }q^{\prime
}\left\langle \hat{q}\right\rangle -4\nu \frac{\Lambda ^{\prime }}{\Lambda }%
p^{\prime \prime }\left\langle \hat{p}\right\rangle \right. \\
+2\mu ^{2}\left[ \left( \frac{\chi ^{\prime }}{\chi }+2\frac{\chi ^{\prime
\prime }}{\chi }q^{\prime 2}\right) +2\left( \frac{\chi ^{\prime }q^{\prime }%
}{\chi }\right) ^{2}\right] \left\langle \hat{q}^{2}\right\rangle +\nu ^{2}%
\left[ \left( \frac{\Lambda ^{\prime }}{\Lambda }+2\frac{\Lambda ^{\prime
\prime }}{\Lambda }p^{\prime \prime 2}\right) +2\left( \frac{\Lambda
^{\prime }p^{\prime \prime }}{\Lambda }\right) ^{2}\right] \left\langle \hat{%
p}^{2}\right\rangle \\
\left. +6\mu \nu \frac{\chi ^{\prime }}{\chi }\frac{\Lambda ^{\prime }}{%
\Lambda }q^{\prime }p^{\prime \prime }\left\langle \hat{q}\hat{p}+\hat{p}%
\hat{q}\right\rangle +\cdots \right\}
\end{gather*}
We therefore obtain the approximate form 
\begin{equation*}
\frac{\mathcal{\hat{V}}\left( q^{\prime },p^{\prime \prime }\right) }{\sqrt{%
\rho \left( q^{\prime },p^{\prime \prime }\right) }}=1-2\mu \frac{\chi
^{\prime }}{\chi }q^{\prime }\left( \hat{q}-\left\langle \hat{q}%
\right\rangle \right) -2\nu \frac{\Lambda ^{\prime }}{\Lambda }p^{\prime
\prime }\left( \hat{p}-\left\langle \hat{p}\right\rangle \right)
\end{equation*}
\begin{eqnarray*}
&&+\mu ^{2}\left[ \left( \frac{\chi ^{\prime }}{\chi }+2\frac{\chi ^{\prime
\prime }}{\chi }q^{\prime 2}\right) \hat{q}^{2}-\left( \frac{\chi ^{\prime }%
}{\chi }+2\frac{\chi ^{\prime \prime }}{\chi }q^{\prime 2}+2\left( \frac{%
\chi ^{\prime }q^{\prime }}{\chi }\right) ^{2}\right) \left\langle \hat{q}%
^{2}\right\rangle \right. \\
&&\left. +6\left( \frac{\chi ^{\prime }q^{\prime }}{\chi }\right)
^{2}\left\langle \hat{q}\right\rangle ^{2}-4\left( \frac{\chi ^{\prime
}q^{\prime }}{\chi }\right) ^{2}\hat{q}\left\langle \hat{q}\right\rangle %
\right] \\
&&+\nu ^{2}\left[ \left( \frac{\Lambda ^{\prime }}{\Lambda }+2\frac{\Lambda
^{\prime \prime }}{\Lambda }p^{\prime \prime 2}\right) \hat{p}^{2}-\left( 
\frac{\Lambda ^{\prime }}{\Lambda }+2\frac{\Lambda ^{\prime \prime }}{%
\Lambda }p^{\prime \prime 2}+2\left( \frac{\Lambda ^{\prime }p^{\prime
\prime }}{\Lambda }\right) ^{2}\right) \left\langle \hat{p}^{2}\right\rangle
\right. \\
&&\left. +6\left( \frac{\Lambda ^{\prime }p^{\prime \prime }}{\Lambda }%
\right) ^{2}\left\langle \hat{p}\right\rangle ^{2}-4\left( \frac{\Lambda
^{\prime }p^{\prime \prime }}{\Lambda }\right) ^{2}\hat{p}\left\langle \hat{p%
}\right\rangle \right] \\
&&+6\mu \nu \frac{\chi ^{\prime }}{\chi }\frac{\Lambda ^{\prime }}{\Lambda }%
q^{\prime }p^{\prime \prime }\left( 2\left( \hat{q}\hat{p}+\hat{p}\hat{q}%
\right) -3\left\langle \hat{q}\hat{p}+\hat{p}\hat{q}\right\rangle
+6\left\langle \hat{q}\right\rangle \left\langle \hat{p}\right\rangle -4\hat{%
q}\left\langle \hat{p}\right\rangle -4\hat{p}\left\langle \hat{q}%
\right\rangle \right) +\cdots .
\end{eqnarray*}

Our objective is to study the random propagator $\frac{\mathcal{\hat{V}}%
\left( Q^{\prime },P^{\prime \prime }\right) }{\sqrt{\rho \left( Q^{\prime
},P^{\prime \prime }\right) }}$ where now the classical random variables $%
Q^{\prime }$ and $P^{\prime \prime }$ have $\rho $ as their joint
probability density function. As we have seen, the exact distributions of $%
Q^{\prime }$ and $P^{\prime \prime }$ are non-trivially related to the
original states of the system and apparatus. However, we are only interested
in the lowest order dependence in terms of parameters $\mu $ and $\nu $. The
decompositions (\ref{decomp}) may be rewritten as

\begin{equation}
Q^{\prime }=Q_{0}^{\prime }+\mu \left\langle \hat{q}\right\rangle +\mu
F,\;P^{\prime \prime }=P_{0}^{\prime \prime }+\nu \left\langle \hat{p}%
\right\rangle +\nu G
\end{equation}
where $Q_{0}^{\prime }$ and $P_{0}^{\prime \prime }$ are independent
variables with the pre-interaction densities $\chi \left( \left|
q_{0}^{\prime }\right| ^{2}\right) ^{2}$ and $\Lambda \left( \left|
p_{0}^{\prime \prime }\right| ^{2}\right) ^{2}$ respectively, and $F$ and $G$
are further independent variables with means of order $\nu $ and $\mu $,
respectively. To lowest order, we obtain 
\begin{eqnarray*}
\frac{\chi ^{\prime }\left( \left| Q^{\prime }\right| ^{2}\right) }{\chi
\left( \left| Q^{\prime }\right| ^{2}\right) }Q^{\prime } &=&\frac{\chi
_{0}^{\prime }}{\chi _{0}}Q_{0}^{\prime }+\mu \left[ \frac{\chi _{0}^{\prime
}}{\chi _{0}}+2\left( \frac{\chi _{0}^{\prime \prime }}{\chi _{0}}-\left( 
\frac{\chi _{0}^{\prime }}{\chi _{0}}\right) ^{2}\right) Q_{0}^{\prime 2}%
\right] \left( \left\langle \hat{q}\right\rangle +F\right) +\cdots \\
\frac{\Lambda ^{\prime }\left( \left| P^{\prime \prime }\right| ^{2}\right) 
}{\Lambda \left( \left| P^{\prime \prime }\right| ^{2}\right) }P^{\prime
\prime } &=&\frac{\Lambda _{0}^{\prime }}{\Lambda _{0}}P_{0}^{\prime \prime
}+\nu \left[ \frac{\Lambda _{0}^{\prime }}{\Lambda _{0}}+2\left( \frac{%
\Lambda _{0}^{\prime \prime }}{\Lambda _{0}}-\left( \frac{\Lambda
_{0}^{\prime }}{\Lambda _{0}}\right) ^{2}\right) P_{0}^{\prime \prime }%
\right] \left( \left\langle \hat{p}\right\rangle +G\right) +\cdots
\end{eqnarray*}
where $\chi _{0}=\chi \left( \left| Q_{0}^{\prime }\right| ^{2}\right)
,\Lambda _{0}=\Lambda \left( \left| P_{0}^{\prime \prime }\right|
^{2}\right) $, etc. Making these replacements leads to 
\begin{equation*}
\frac{\mathcal{\hat{V}}\left( Q^{\prime },P^{\prime \prime }\right) }{\sqrt{%
\rho \left( Q^{\prime },P^{\prime \prime }\right) }}=1-2\mu \frac{\chi
_{0}^{\prime }}{\chi _{0}}Q_{0}^{\prime }\left( \hat{q}-\left\langle \hat{q}%
\right\rangle \right) -2\nu \frac{\Lambda _{0}^{\prime }}{\Lambda _{0}}%
P_{0}^{\prime \prime }\left( \hat{p}-\left\langle \hat{p}\right\rangle
\right)
\end{equation*}
\begin{eqnarray}
&&+\mu ^{2}\left[ \left( \frac{\chi _{0}^{\prime }}{\chi _{0}}+2\frac{\chi
_{0}^{\prime \prime }}{\chi _{0}}Q_{0}^{\prime 2}\right) \hat{q}^{2}-\left( 
\frac{\chi _{0}^{\prime }}{\chi _{0}}+2\frac{\chi _{0}^{\prime \prime }}{%
\chi _{0}}Q_{0}^{\prime 2}+2\left( \frac{\chi _{0}^{\prime }Q_{0}^{\prime }}{%
\chi _{0}}\right) ^{2}\right) \left\langle \hat{q}^{2}\right\rangle +6\left( 
\frac{\chi _{0}^{\prime }Q_{0}^{\prime }}{\chi _{0}}\right) ^{2}\left\langle 
\hat{q}\right\rangle ^{2}\right.  \notag \\
&&\left. -4\left( \frac{\chi _{0}^{\prime }Q_{0}^{\prime }}{\chi _{0}}%
\right) ^{2}\hat{q}\left\langle \hat{q}\right\rangle -2\left[ \frac{\chi
_{0}^{\prime }}{\chi _{0}}+2\left( \frac{\chi _{0}^{\prime \prime }}{\chi
_{0}}-\left( \frac{\chi _{0}^{\prime }}{\chi _{0}}\right) ^{2}\right)
Q_{0}^{\prime 2}\right] \left( \left\langle \hat{q}\right\rangle +F\right)
\left( \hat{q}-\left\langle \hat{q}\right\rangle \right) \right]  \notag \\
&&+\nu ^{2}\left[ \left( \frac{\Lambda _{0}^{\prime }}{\Lambda _{0}}+2\frac{%
\Lambda _{0}^{\prime \prime }}{\Lambda _{0}}P_{0}^{\prime \prime 2}\right) 
\hat{p}^{2}-\left( \frac{\Lambda _{0}^{\prime }}{\Lambda _{0}}+2\frac{%
\Lambda _{0}^{\prime \prime }}{\Lambda _{0}}P_{0}^{\prime \prime 2}+2\left( 
\frac{\Lambda _{0}^{\prime }P_{0}^{\prime \prime }}{\Lambda _{0}}\right)
^{2}\right) \left\langle \hat{p}^{2}\right\rangle +6\left( \frac{\Lambda
_{0}^{\prime }P_{0}^{\prime \prime }}{\Lambda _{0}}\right) ^{2}\left\langle 
\hat{p}\right\rangle ^{2}\right.  \notag \\
&&\left. -4\left( \frac{\Lambda _{0}^{\prime }P_{0}^{\prime \prime }}{%
\Lambda _{0}}\right) ^{2}\hat{p}\left\langle \hat{p}\right\rangle -2\left[ 
\frac{\Lambda _{0}^{\prime }}{\Lambda _{0}}+2\left( \frac{\Lambda
_{0}^{\prime \prime }}{\Lambda _{0}}-\left( \frac{\Lambda _{0}^{\prime }}{%
\Lambda _{0}}\right) ^{2}\right) P_{0}^{\prime \prime }\right] \left(
\left\langle \hat{p}\right\rangle +G\right) \left( \hat{p}-\left\langle \hat{%
p}\right\rangle \right) \right]  \notag \\
&&+6\mu \nu \frac{\chi _{0}^{\prime }}{\chi _{0}}\frac{\Lambda _{0}^{\prime }%
}{\Lambda _{0}}Q_{0}^{\prime }P_{0}^{\prime \prime }\left( 2\left( \hat{q}%
\hat{p}+\hat{p}\hat{q}\right) -3\left\langle \hat{q}\hat{p}+\hat{p}\hat{q}%
\right\rangle +6\left\langle \hat{q}\right\rangle \left\langle \hat{p}%
\right\rangle -4\hat{q}\left\langle \hat{p}\right\rangle -4\hat{p}%
\left\langle \hat{q}\right\rangle \right) +\cdots .  \label{expansion}
\end{eqnarray}

\section{Repeated Measurements}

Next suppose that simultaneous measurements of position and momentum are
made at regular intervals of time $\tau $. Again, we assume that the
pre-interaction states $\phi _{A\left( j\right) }$ are all copies of the
fixed state $\phi _{A}$.

If we denote the conditioned wave-function immediately after the $n$-th
measurement by $\psi _{n}=\psi _{n}\left( x|Q_{1}^{\prime },P_{1}^{\prime
\prime };\cdots ,Q_{n}^{\prime },P_{n}^{\prime \prime }\right) $, then we
have the relation 
\begin{equation*}
\psi _{n}=e^{\tau H/i\hbar }\;\frac{\mathcal{\hat{V}}\left( Q_{n}^{\prime
},P_{n}^{\prime \prime }\right) }{\sqrt{\rho \left( Q_{n}^{\prime
},P_{n}^{\prime \prime }\right) }}\psi _{n-1}.
\end{equation*}

\bigskip

Setting $\mu =\nu =\sqrt{\tau }$, we can use (\ref{expansion}) to obtain 
\begin{gather}
\frac{1}{\tau }[\psi _{n}-\psi _{n-1}]=\left\{ \frac{1}{i\hbar }\hat{H}-2%
\sqrt{\frac{1}{\tau }}\left( \frac{\chi _{0}^{\prime }}{\chi _{0}}%
Q_{0}^{\prime }\right) \left( q-\left\langle \hat{q}\right\rangle \right) -2%
\sqrt{\frac{1}{\tau }}\left( \frac{\Lambda _{0}^{\prime }}{\Lambda _{0}}%
P_{0}^{\prime \prime }\right) \left( \hat{p}-\left\langle \hat{p}%
\right\rangle \right) \right.  \notag \\
\left. -2\frac{1}{\tau }\left( \frac{\chi _{0}^{\prime }}{\chi _{0}}%
Q_{0}^{\prime }\right) _{n}^{2}\left( q-\left\langle \hat{q}\right\rangle
\right) ^{2}-2\frac{1}{\tau }\left( \frac{\Lambda _{0}^{\prime }}{\Lambda
_{0}}P_{0}^{\prime \prime }\right) _{n}^{2}\left( \hat{p}-\left\langle \hat{p%
}\right\rangle \right) ^{2}\right\} \psi _{n-1}+O\left( \tau ^{1/2}\right) .
\label{approx}
\end{gather}
To obtain (\ref{approx}) from (\ref{expansion}) we made the replacements $%
\frac{\chi _{0}^{\prime }}{\chi _{0}}+2\frac{\chi _{0}^{\prime \prime }}{%
\chi _{0}}Q_{0}^{\prime 2}$ with $-2\left( \frac{\chi _{0}^{\prime }}{\chi
_{0}}Q_{0}^{\prime }\right) ^{2}$ and $\frac{\Lambda _{0}^{\prime }}{\Lambda
_{0}}+2\frac{\Lambda _{0}^{\prime \prime }}{\Lambda _{0}}P_{0}^{\prime
\prime 2}$ with $-2\left( \frac{\Lambda _{0}^{\prime }}{\Lambda _{0}}%
P_{0}^{\prime \prime }\right) ^{2}$ which is consistent with (\ref{kappa}).
In particular, the coefficients of $\left\langle \hat{q}^{2}\right\rangle $
and $\left\langle \hat{p}^{2}\right\rangle $ disappear. The terms involving $%
F$ and $G$ are dropped since they are $O\left( \tau ^{1/2}\right) $ and so
will give negligible contribution in a law of large numbers limit. Finally
the cross terms vanish since 
\begin{equation*}
\lim_{\tau \rightarrow 0}\frac{1}{\tau }\sum_{j=1}^{\left[ t/\tau \right]
}\left( \frac{\chi _{0}^{\prime }}{\chi _{0}}Q_{0}^{\prime }\right)
_{j}\left( \frac{\Lambda _{0}^{\prime }}{\Lambda _{0}}P_{0}^{\prime \prime
}\right) _{j}=0
\end{equation*}
on account of the fact that $\int \chi \left( y^{2}\right) \chi ^{\prime
}\left( y^{2}\right) ydy=0=\int \Lambda \left( x^{2}\right) \Lambda ^{\prime
}\left( x^{2}\right) xdx$.

\bigskip

Comparison with position-only situation shows that we are lead to the
stochastic differential equation 
\begin{eqnarray*}
\left| d\Psi \right\rangle &=&\left\{ \frac{1}{i\hbar }\hat{H}-\kappa
_{q}\left( \hat{q}-\left\langle \hat{q}\right\rangle \right) ^{2}-\kappa
_{p}\left( \hat{p}-\left\langle \hat{p}\right\rangle \right) ^{2}\right\}
\left| \Psi \right\rangle \,dt \\
&&-\sqrt{2\kappa _{q}}\left( \hat{q}-\left\langle \hat{q}\right\rangle
\right) \left| \Psi \right\rangle \,dB_{t}^{\left( q\right) }-\sqrt{2\kappa
_{p}}\left( \hat{p}-\left\langle \hat{p}\right\rangle \right) \left| \Psi
\right\rangle \,dB_{t}^{\left( p\right) }
\end{eqnarray*}
where 
\begin{equation*}
\kappa _{q}=2\int \left( \chi ^{\prime }\left( y^{2}\right) y\right)
^{2}dy,\quad \kappa _{p}=2\int \left( \Lambda ^{\prime }\left( x^{2}\right)
x\right) ^{2}dx
\end{equation*}
and we have the\ following limits in mean-square sense to independent Wiener
processes 
\begin{equation*}
B_{t}^{\left( q\right) }=\lim_{\tau \rightarrow 0}\sqrt{\frac{2}{\kappa
_{q}\tau }}\sum_{j=1}^{\left[ t/\tau \right] }\left( \frac{\chi _{0}^{\prime
}}{\chi _{0}}Q_{0}^{\prime }\right) _{j},\quad B_{t}^{\left( p\right)
}=\lim_{\tau \rightarrow 0}\sqrt{\frac{2}{\kappa _{p}\tau }}\sum_{j=1}^{%
\left[ t/\tau \right] }\left( \frac{\Lambda _{0}^{\prime }}{\Lambda _{0}}%
P_{0}^{\prime \prime }\right) _{j}.
\end{equation*}

\subsection{Stochastic Dynamics}

The associated Lindblad generator for this monitored dynamics is 
\begin{equation*}
\mathcal{L}\left( \hat{Y}\right) =\frac{1}{i\hbar }\left[ \hat{Y},\hat{H}%
\right] +\kappa _{q}\left\{ \left[ \hat{q},\hat{Y}\right] \hat{q}+\hat{q}%
\left[ \hat{Y},\hat{q}\right] \right\} +\kappa _{p}\left\{ \left[ \hat{p},%
\hat{Y}\right] \hat{p}+\hat{p}\left[ \hat{Y},\hat{p}\right] \right\} .
\end{equation*}
We note that $\mathcal{L}\left( \hat{Y}\right) =\frac{1}{i\hbar }\left[ \hat{%
Y},\hat{H}\right] $occurs for the special case of observables of the type $%
\hat{Y}=\alpha \hat{q}+\beta \hat{p}+\gamma \hat{q}\hat{p}$. This means that
when $\hat{H}=\frac{1}{2m}\hat{p}^{2}+\Phi \left( \hat{q}\right) $,\ the
averages of the canonical observables evolve in a non-random way according
to the Ehrenfest theorem for closed systems: 
\begin{equation*}
\frac{d}{dt}\left\langle \hat{q}\right\rangle =\frac{1}{m}\left\langle \hat{p%
}\right\rangle ,\quad \frac{d}{dt}\left\langle \hat{p}\right\rangle
=-\left\langle \Phi ^{\prime }\left( \hat{q}\right) \right\rangle .
\end{equation*}
However, we do not have the derivational property $\mathcal{L}\left( \hat{X}%
\hat{Y}\right) =\hat{X}\mathcal{L}\left( \hat{Y}\right) +\mathcal{L}\left( 
\hat{X}\right) \hat{Y}$ and so the dynamics is dissipative. In particular,

\begin{equation*}
\mathcal{L}\left( \hat{H}\right) =\kappa _{q}\frac{\hbar ^{2}}{m}+\kappa
_{p}\hbar ^{2}\Phi ^{\prime \prime }\left( \hat{q}\right)
\end{equation*}
and so energy is not conserved.

\subsection{Conclusion}

We have shown that the central limit effect allows us to derive a stochastic
Schr\"{o}dinger equation in a very general setting. Essentially we need only
be in the domain of attraction for Gaussian statistics. (It is possible that
more general results hold for stable laws - leading to stochastic
Schr\"{o}dinger equations driven by Levy processes - however, this clearly
would be indicative of an imperfection in the measurement apparatus only.)
The striking result is that the non-commuting observables can be measured
simultaneously with negligible interference. This backs up the general
phenomenological approach used in\ areas of quantum control and filtering.
In their article, Scott \& Milburn \cite{SM} investigate several models,
including classically chaotic ones, and show that the system is localized
with the monitored trajectory in phase space corresponding to a quantum
average plus noise. Whereas there are still many interesting open questions
regarding, for instance, the semi-classical limit, the present result does
show claim that the analysis of Scott \& Milburn approach is generic for
continuously monitored phase variables.

\section{Appendix}

Let $\left( q,p\right) \in \Gamma $ where $\Gamma =\mathbb{R}^{2}$ is phase
space. For a given function $f=f\left( q,p\right) $ on phase space, the
association of an operator $f\left( \hat{q},\hat{p}\right) $ is ambiguous
due to the problem of operator ordering. We shall adopted the Weyl
quantization convention.

The Weyl operator at phase point $\left( q,p\right) $ is defined to be $\hat{%
W}\left( q,p\right) =\exp \left( q\hat{p}-p\hat{q}\right) /i\hbar $. From
the CCR and an application of the Baker-Hausdorff-Campbell theorem, we find
that 
\begin{equation}
\hat{W}\left( q,p\right) \hat{W}\left( q^{\prime },p^{\prime }\right)
=e^{\left( qp^{\prime }-pq^{\prime }\right) /2i\hbar }\hat{W}\left(
q+q^{\prime },p+p^{\prime }\right) .
\end{equation}

Let $f=f\left( q,p\right) $ be absolutely integrable on $\Gamma $ and define
its phase space Fourier (Weyl) transform to be 
\begin{equation}
\tilde{f}\left( q,p\right) :=\int_{\Gamma }\frac{d\bar{q}d\bar{p}}{2\pi
\hbar }\,e^{\left( p\bar{q}-q\bar{p}\right) /i\hbar }\,f\left( \bar{q},\bar{p%
}\right) .
\end{equation}
$\allowbreak $The Weyl quantization of $f$ is then defined to be the
operator 
\begin{eqnarray}
\left[ f\left( \hat{q},\hat{p}\right) \right] _{\text{Weyl}}
&:&=\int_{\Gamma }\frac{dqdp}{2\pi \hbar }\;\hat{W}\left( q,p\right) \tilde{f%
}\left( q,p\right)  \notag \\
&=&\int_{\Gamma \times \Gamma }\frac{dqdp}{2\pi \hbar }\frac{d\bar{q}d\bar{p}%
}{2\pi \hbar }\,e^{\left( p\bar{q}-q\bar{p}\right) /i\hbar }\,f\left( \bar{q}%
,\bar{p}\right) \;\hat{W}\left( q,p\right)
\end{eqnarray}
and we refer to the map $f\left( q,p\right) \mapsto \left[ f\left( \hat{q},%
\hat{p}\right) \right] _{\text{Weyl}}$ as Weyl quantization.

We have for instance $\left[ \exp \left( t\left( q\hat{p}-p\hat{q}\right)
/i\hbar \right) \right] _{\text{Weyl}}=\hat{W}\left( tq,tp\right) $ and
expanding in powers of $t$, we obtain $\left[ \left( \hat{q}\alpha +\hat{p}%
\beta \right) ^{n}\right] _{\text{Weyl}}=\left( \hat{q}\alpha +\hat{p}\beta
\right) ^{n}$. Further expansion in terms of $\alpha ,\beta $ show that
polynomials will be mapped to the symmetrically (Weyl) ordered form.. For
instance, $\left[ \hat{q}^{2}\hat{p}\right] _{\text{Weyl}}=\frac{1}{3}\left( 
\hat{q}^{2}\hat{p}+\hat{q}\hat{p}\hat{q}+\hat{p}\hat{q}^{2}\right) $, etc.

\begin{acknowledgement}
We acknowledge very useful conversations with Professors Smolianov and
Truman. We also wish to thank Professor Daniel Heffernan for valuable
insights into the problem quantum measurement of phase space variables and
their relation to quantum chaos.
\end{acknowledgement}

\end{document}